\documentclass[12pt]{iopart}

\usepackage{iopams}  
\usepackage[dvipsnames]{xcolor}

\usepackage{graphicx}

\begin{document}
\title[Isotopic substitution in ultracold molecule reactions]{Product-state distribution after  isotopic substitution in ultracold atom-molecule collisions}

\author{Maciej B. Kosicki,  Piotr S. \.{Z}uchowski}
 \ead{pzuch@fizyka.umk.pl}
 \address{Institute of Physics, Faculty of Physics, Astronomy and Informatics, Nicolaus Copernicus University, Grudziadzka 5, 87-100 Toru\'{n}, Poland}

\author{Maykel L. Gonz\'{a}lez-Mart\'{i}nez and Olivier Dulieu}%
\ead{olivier.dulieu@u-psud.fr}
\address{%
Laboratoire Aim\'{e} Cotton, CNRS, Universit\'{e} Paris-Sud, ENS Paris-Saclay, Universit\'e Paris-Saclay, \\
B\^{a}t. 505, Campus d’Orsay, 91405 Orsay, France
}%

\vspace{10pt}
\begin{indented}
\item[]\today     
\end{indented}

\begin{abstract}
We show that products of the isotopic substitution reactions in experimentally accessible molecules such as NaK, RbCs and SrF are cold according to their translational energy below hundreds of mK. For these chemical reactions molecular products may occupy only the lowest rotational states. We also discuss the possibility of controlling the chemical reactions by the electric field in ultracold mixtures of molecules and atoms with low kinetic energy release, where one of the constituent atoms of colliding molecule is replaced by its isotope. This letter open new avenues in investigating the branching ratios of chemical reactions in ultracold conditions. 
\end{abstract}

\submitto{\jpb}
\maketitle

The past decade has witnessed rapid progress in the investigation of chemistry at ultracold temperatures. 
Ospelkaus \textit{et al.}~\cite{Ospelkaus:2009} demonstrated field-free chemical reactions between KRb molecules and constituent atoms. 
By controlling the internal state of KRb molecule, it was possible to enhance the reaction rate between two KRb molecules by allowing $s$-wave collisions. They have also described collisions of KRb with ultracold K or Rb atoms finding that a formation of K+Rb${_2}$ products is energetically forbidden. Later on, the same experimental group~\cite{Ni:2010,deMiranda:2011} has also shown that an external electric field may be applied to orient KRb molecules and to influence their reaction rate. Recently, a mechanism to monitor the chemical reactivity by the selection of reactant vibrational state has been utilized in the reaction between NaRb molecules \cite{Ye:2018}. 
Other experimental attempts involve Feshbach molecules and ultracold atoms in the presence of an external magnetic field~\cite{Knoop:2010,Rui:2017}.

Measurements of the reactant loss is a straightforward way to  probe reactivity at ultracold temperatures. However, the detection of products would provide much more details about the underlying dynamics. A basic difficulty in detecting product states of such reactions is that the kinetic energy release is often larger by several orders of magnitude than the depths of most traps used in experiments (see \textit{e.g.} \cite{Zuchowski:2010,Kosicki:2017}). Despite several successful realizations of ultracold molecular gases of ground-state heteronuclear alkali-metal diatomic species \cite{Zwierlein:2015,Will:2016,Wang:2016,Cornish:2014,Takekoshi:2014,Truppe:2017,Williams:2018,McCarron:2018,Anderegg:2018} and promising prospects \cite{Borsalino:2016, Stevenson:2016}, no detailed study of product-state distribution has yet been reported so far. However two important steps in this direction were recently achieved: the detection of weakly-bound $^{87}$Rb${_2}$ molecules by three-Rb-atom recombination \cite{Harter:2013, Wolf:2017}, and the detection of polyatomic ions after the collision of two ultracold KRb molecules \cite{Hu:2019}. 

It would be desirable to explore cases in which the products can be accumulated in the trap after the chemical reaction. Inspired by a recent proposal for chemical reaction with very small energy release, involving isotope exchange in ultracold molecular collisions \cite{Tomza:2015:b}, we investigate in the present work the isotopic substitution reactions between a ground-state alkali-metal atom $^a$A with atomic mass $a$ and a ground-state heteronuclear diatomic molecule containing a different isotopic species $^b$A (with $a>b$), and another atom B
\begin{equation}
{^a}\mathrm{A} + {^b}\mathrm{A}\mathrm{B} \longrightarrow {^b}\mathrm{A}+{^a}\mathrm{A}\mathrm{B}.
\label{eq:iso-ex}
\end{equation}
If the molecule is initially in its lowest rovibrational level $v=0, n=0$, (neglecting the hyperfine structure for now), the energy release $\Delta  E_{a-b}$ for the reaction (\ref{eq:iso-ex}) can be expressed in terms of the change of the zero-point energy
\begin{equation}
\Delta  E_{a-b} = \frac{\omega_{a} - \omega_{b}}{2},
\label{eq:deltaE}
\end{equation}
where $\omega_{a}$ ($\omega_{b}$) is the harmonic constant of the ${^a}$AB (${^b}$AB) molecule. Values of $\Delta E_{a-b}$ for a selected series of species of experimental interest are found small enough to be considered in future experiments based for instance on electrostatic or microwave traps (Table~\ref{tbl:zero_field}).

Our previous studies show that atom-exchange reactions involving alkali-metal dimers and atoms, as well as strontium monofluoride and strontium atoms, are barrierless processes~\cite{Zuchowski:2010,Kosicki:2017}. In all these cases the potential energy surfaces (PES) are very deep and support tens of vibrational levels and hundreds of rotational levels. Full quantum scattering calculations for atom-molecule reactive collisions, even for the field-free case, are cumbersome for strongly interacting species as alkali-metal atoms~\cite{Hutson:2006,Cvitas:2005a,Cvitas:2005b,Cvitas:2007,Quemener:2005}, due to the huge number of involved channels. Interestingly enough, the isotope-exchange of lithium in field-free atom-diatom collisions has been investigated in~\cite{Cvitas:2005b}, and the product-state distribution for this system has been analyzed.

Recently, it was demonstrated~\cite{Croft:2017a,Croft:2017b} that the distribution of product rotational states after the atom-exchange reaction between ultracold K atom and KRb molecule, calculated with rigorous quantum scattering calculations exhibits the Poissonian  distribution, while  the of positions and widths of scattering resonances display signatures of quantum chaos: Wigner-Dyson distribution of resonances spacings in energy domain, and  Porter-Thomas distribution of widths of resonances.
A similar conclusion has been reached for the ${\rm D}^++{\rm H}_2\to{\rm H}^+ +{\rm DH}$ reaction~\cite{Lara:2015a, Lara:2015b}. Furthermore, it was noticed in \cite{Mayle:2012} that the collisions of alkali-metal dimers with atoms exhibit strongly resonant character: near the collision threshold, the density of bound states supported by closed channels (related to excited hyperfine, rotational and vibrational states of products) is very high. Because of that, long-lived collision complexes are predicted to be formed, and full randomization of the reaction energy (via internal modes) should occur before the complex is destroyed. This essentially makes the product state distribution of the complex \textit{  statistical, i.e.} without featuring correlation between products and reactants. We then use the statistical approach adopted in~\cite{Statfields:2014} for ultracold molecular collisions to investigate such isotope-substitution reactions when all particles are in their absolute ground state in free space. We will show how the electric field can be used to control their reactivities by blocking the reaction channels and by tuning product state distributions. Although we focus here on the electric field control, one could imagine a similar scenario with the magnetic field for paramagnetic molecules or microwave radiation fields.   

For the chemical reactions addressed in Table~\ref{tbl:zero_field}, the kinetic energy release is small. Values for $\Delta  E_{a-b}$ values have been obtained using experimental data~\cite{Ivanova:2011,Jastrzebski:2008,Ross:1990,Ferber:2008,Kasahara:1996,Fellows:1999,Bernath:1996} and mass-scaled harmonic frequencies of the ground state. Non-Born-Oppenheimer (NBO) effects, leading to an estimated shift of about dozens of~MHz between the two isotopologues ${^a}$AB and ${^b}$AB for alkali-metal dimers~\cite{Lutz:2016}, are neglected, so that they have both the same dissociation limit before including hyperfine structure. The same origin is considered for the hyperfine manifold of the colliding atomic partner. The values of $\Delta E_{a-b}$ are obtained after setting the origin of energies at this dissociation limit, which is the center of gravity of the hyperfine manifold of the constituent atoms. The hyperfine structure of the $(v=0, n=0)$ ground-state level is also neglected for these evaluations. The $^{88}$Sr+$^{86}$Sr$^{19}$F reaction must be considered with care, as $\Delta E_{a-b}$ is only about 300 MHz larger than the $n=0 \to n=1$ rotational excitation in $^{88}$Sr$^{19}$F. Using the same methodology as in~\cite{Lutz:2016}, based on ADF quantum-chemistry code~\cite{ADF2001} with the density functional method (PBE96)~\cite{PBE96}, we found that the NBO shift between the ground-state $(v=0, n=0)$ level energies of $^{88}$Sr$^{19}$F and $^{86}$Sr$^{19}$F, is smaller than 10~MHz, and thus, can be safely neglected.

\begin{table*}[ht]
\caption{The calculated reaction energy $\Delta E_{a-b}$ (references to used data in square brackets),  and the rotational constant $B_0$, the hyperfine coupling constant $A$ and spin-orbit coupling constant $\gamma_{SR}$ of products are shown. Symbols $n_{\rm max}$ and ${\ell}_{\rm max}$ denote the maximum  
rotational state of molecule for a collision energy of 1 $\mu$K and  the maximum end-over-end angular momentum which can be populated by $\Delta E_{a-b}$, respectively. See supplementary material for complete compilation for all possible alkali-metal dimers reactions.}
\centering
\label{tbl:zero_field}
\begin{tabular}{p{0.2\linewidth}p{0.14\linewidth}p{0.15\linewidth}p{0.14\linewidth}p{0.12\linewidth}p{0.05\linewidth}p{0.05\linewidth}p{0.05\linewidth}}
\hline
Products & $\Delta E_{a-b}$  $[$mK$]$ &   $B_0/{k_b T}$ $[$mK$]$ & $A/{k_b T}$ $[$mK$]$& $\gamma_{SR}$ $[$mK$]$ & $n_{\rm max}$ &  ${\ell}_{\rm max}$\\
\hline
${}^{39}$K $+$ ${}^{23}$Na${}^{40}$K		 	&  415~\cite{Jastrzebski:2008}  	& 136~\cite{Ross:2000}	 		& 11~\cite{Arimondo:1977}  & - & 1 & 18 \\
${}^{39}$K $+$ ${}^{23}$Na${}^{41}$K 		& 811~\cite{Jastrzebski:2008} 	&  136~\cite{Ross:2000} 		& 11~\cite{Arimondo:1977}& - & 1 & 23\\
${}^{40}$K $+$ ${}^{23}$Na${}^{41}$K 		&  396~\cite{Jastrzebski:2008} 	& 136~\cite{Ross:2000} 		& -14~\cite{Arimondo:1977}& - &   1 & 18\\
${}^{39}$K $+$ ${}^{40}$K${}^{87}$Rb	 		& 472~\cite{Ross:1990} 				& 55~\cite{Ni:2008} 				& 11~\cite{Arimondo:1977}& - & 2 & 22 \\
${}^{39}$K $+$ ${}^{40}$K${}^{133}$Cs 		& 478~\cite{Ferber:2008}			& 44~\cite{Deiglmayr:2008} 	& 11~\cite{Arimondo:1977} & - & 2 & 23 \\
${}^{85}$Rb $+$ ${}^{7}$Li${}^{87}$Rb 		& 123~\cite{Ivanova:2011} 			&  316~\cite{Deiglmayr:2008} 	& 49~\cite{Bize:1999} & - &  0 & 16 \\
${}^{85}$Rb $+$ ${}^{23}$Na${}^{87}$Rb 	& 188~\cite{Kasahara:1996} 		&  100~\cite{Wang:2016} 	 	& 49~\cite{Bize:1999}  & - & 0 & 20 \\
${}^{85}$Rb $+$ ${}^{39}$K${}^{87}$Rb 		& 198~\cite{Ross:1990}				&  55~\cite{Ni:2008} 				& 49~\cite{Bize:1999}  & - & 1 & 22 \\
${}^{85}$Rb $+$ ${}^{87}$Rb${}^{133}$Cs 	& 253~\cite{Fellows:1999}	 		& 23~\cite{Childs:1981} 		& 49~\cite{Bize:1999}  & - & 1 & 27 \\
${}^{86}$Sr $+$ ${}^{88}$Sr${}^{19}$F 		& 745~\cite{Bernath:1996}			& 365~\cite{Bernath:1996}	 &  5~\cite{Childs:1981}  &  4~\cite{Childs:1981}  & 1 & 27 \\
\hline
\end{tabular}
\end{table*}

The $\Delta E_{a-b}$ values displayed in Table~\ref{tbl:zero_field} range from 123~mk to 811~mK, corresponding to the $^{87}$Rb+$^{7}$Li$^{85}$Rb and $^{41}$K+$^{23}$Na$^{39}$K collisions, respectively. These energy releases are larger than the typical depth of optical dipole traps, but it should be possible to retain the products in electrostatic traps or in microwave cavities, or even in magnetic traps for the case of monofluoride reactants. The maximal value $n_{\rm max}$ of the molecular rotational quantum number $n$ allowed by the energy release is deduced within the rigid-rotor approximation for the rotational energies $E(n)=n(n+1)B_0$, where $B_0$ is the rotational constant of the $(v=0, n=0)$ ground-state level. We neglected collision energy of 1 $\mu$K in the $n_{\rm max}$ estimation. Interestingly, very few rotational levels can be populated in the products, and only $n=0$ in the cases of $^{87}$Rb+$^{85}$Rb$^{23}$Na, and $^{87}$Rb+$^{85}$Rb$^{7}$Li collisions.
In contrast, very large number of hyperfine levels can be  populated in the alkali-metal.  Specifically, there are 144 hyperfine levels for both $n=0$ and $n=1$ rotational states of the Na$^{40}$K and even more for the ${}^{87}$RbCs due to the presence of $n=2$ rotational state (both in X${}^1\Sigma^+$ ground-state). For the alkali-metal-atom$+$ alkali-metal-dimer system, these levels are additionally split by the hyperfine interactions of the atomic collision partner. The rotational splitting in the SrF molecule in the X${}^1\Sigma^-$ ground-state needs additional comments. Firstly, due to non-zero electronic spin, each rotational level is split by the spin-rotation interaction into states described by $j = n + s$. Secondly, these states are split by the fluorine hyperfine interaction. Overall,  there are 16 hyperfine levels in the ${}^{88}$SrF which can be populated for $n = 0$ and $n = 1$ rotational states. In field free case we use $f=j+i$ quantum number to label them, accordingly.

\begin{figure*}[!tbp]
  \centering
  \begin{minipage}[b]{\textwidth}
    \includegraphics[width=\textwidth]{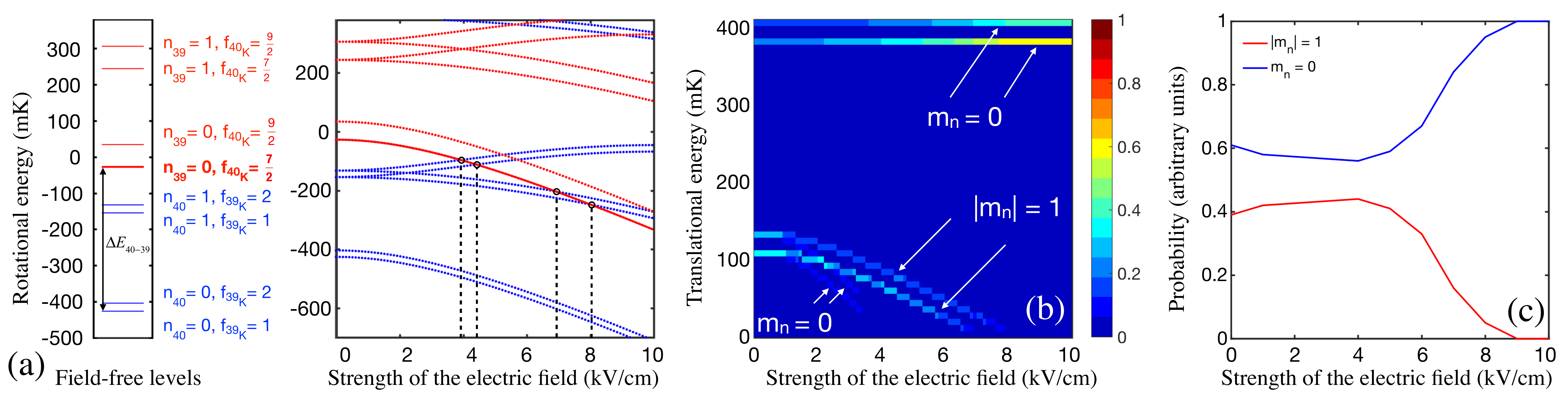}
   \caption{The ${}^{40}$K + $^{23}$Na${}^{39}$K $\longrightarrow$ ${}^{39}$K + ${}^{23}$Na${}^{40}$K reaction. (a) Splitting of the molecular rotational levels in a absence and presence of external electric field. The diagram shows field-free rotational levels which are labeled by $n_{39}$ (indicating $^{23}$Na${}^{39}$K) and $n_{40}$ (indicating ${}^{23}$Na${}^{40}$K) quantum numbers as well as by $f=i+s$ quantum number describing the hyperfine interactions of the atomic partner(${}^{40}$K for reactants and ${}^{39}$K for products). The plot shows a splitting of these levels due to the interaction with external electric field. Note that, the hyperfine states of the system are superimposed the resolution of the diagram and plot. The $\Delta E_{40-39}$ is the reaction exothermicity. The red thick line corresponds to initial state of the reactants. The dashed vertical lines in the plot indicate crossings between reactant and product states due to the Stark effect. (b)  Product distribution of the translational energy. (c) Distribution of the projection of the rotational quantum number, $m_n$,  corresponding to ${}^{23}$Na${}^{40}$K molecule.  The results presented in panels (b) and (c)  are obtained with an initial collisional energy of 1 $\mu$K. } 
    \label{fig1}
  \end{minipage}
  \end{figure*}
\begin{figure*}[!tbp]
  \centering
  \begin{minipage}[b]{\textwidth}
    \includegraphics[width=\textwidth]{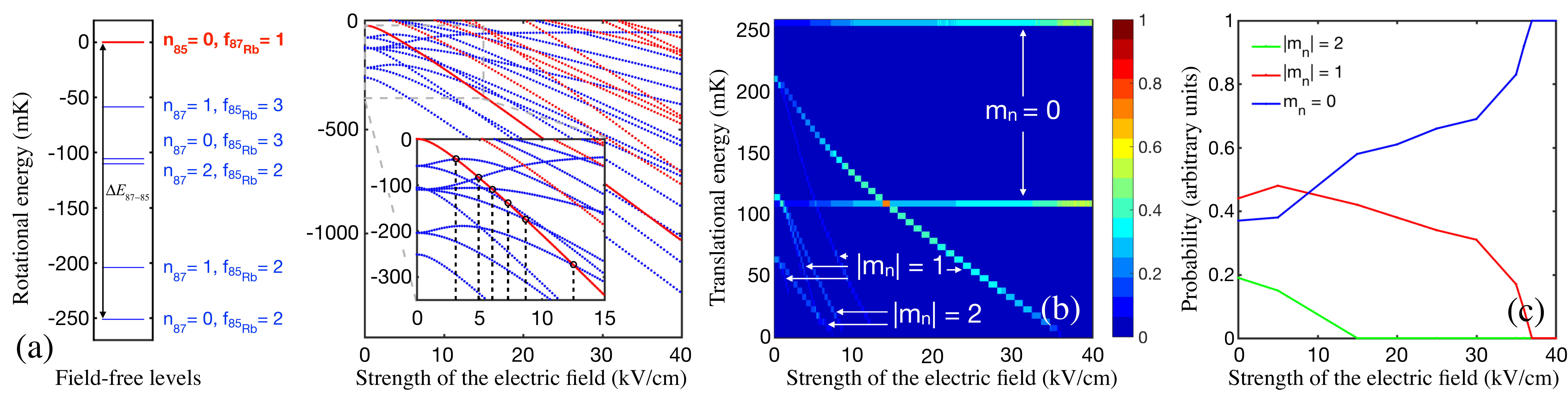}
   \caption{The ${}^{87}$Rb + ${}^{85}$Rb${}^{133}$Cs $ \longrightarrow$ ${}^{85}$Rb + ${}^{87}$Rb${}^{133}$Cs reaction. (a) Splitting of the molecular rotational levels in a absence and presence of external electric field. The diagram shows field-free rotational levels which are labeled by $n_{85}$ (indicating ${}^{85}$Rb${}^{133}$Cs) and $n_{87}$ (indicating ${}^{87}$Rb${}^{133}$Cs) quantum numbers as well as by $f=i+s$ quantum number describing the hyperfine interactions of the atomic partner(${}^{87}$Rb for reactants and ${}^{85}$Rb for products). The plot shows a splitting of these levels due to the interaction with external electric field. Note that, the hyperfine states of the system are superimposed the resolution of the diagram and plot. The $\Delta E_{87-85}$ is the reaction exothermicity. The red thick line corresponds to initial state of the reactants. The dashed vertical lines in the plot indicate crossings between reactant and product states due to the Stark effect. (b)  Product distribution of the translational energy. (c) Distribution of the projection of the rotational quantum number, $m_n$,  corresponding to ${}^{87}$Rb${}^{133}$Cs molecule. The results presented in panels (b) and (c)  are obtained with an initial collisional energy of 1 $\mu$K.  }
    \label{fig2}
  \end{minipage}
\end{figure*}
\begin{figure*}[!tbp]
  \centering
  \begin{minipage}[b]{\textwidth}
    \includegraphics[width=\textwidth]{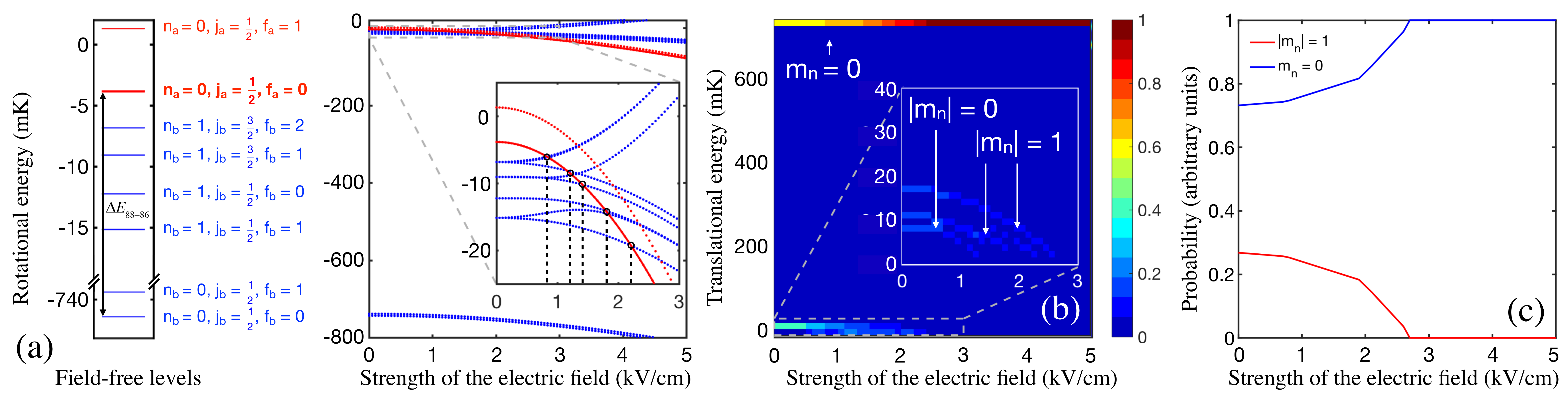}
   \caption{The ${}^{88}$Sr + ${}^{86}$Sr${}^{19}$F $ \longrightarrow$ ${}^{86}$Sr + ${}^{88}$Sr${}^{19}$F reaction. (a) Splitting of the molecular rotational levels in a absence and presence of external electric field. The diagram shows field-free rotational levels which are labeled by the $n$, $j=n+s$, and $f=n+s+i$ quantum numbers which characterize to rotation, spin-rotation and hyperfine couplings in the SrF molecule. The subscripts $a$ and $b$ indicates reactants and products, respectively. The plot shows a splitting of these levels due to the interaction with external electric field. Note that, the hyperfine states of the system are superimposed the resolution of the diagram and plot. The $\Delta E_{a-b}$ is the reaction exothermicity. The red thick line corresponds to initial state of the reactants. The dashed vertical lines in the plot indicate crossings between reactant and product states due to the Stark effect. (b)  Product distribution of the translational energy. (c) Distribution of the projection of the rotational quantum number, $m_n$,  corresponding to ${}^{88}$Sr${}^{19}$F  molecule. The results presented in panels (b) and (c)  are obtained with an initial collisional energy of 1 $\mu$K. }
    \label{fig3}
  \end{minipage}
\end{figure*}

Cold polar molecules considered in this paper exhibit a permanent dipole moment (PDM) in their own frame (identical for all isotopologues to a very good approximation) large enough (see, \textit{e.g.} \cite{Aymar:2005}) to strongly interact with a static electric field. Noticeable Stark shifts comparable to $\Delta E_{a-b}$ at zero field can be induced by applying an experimentally achievable electric field. Figures \ref{fig1}(a), \ref{fig2}(a), and \ref{fig3}(a) show the variation of the level energies of the reactant and product complexes with the electric field for three typical cases, $^{40}$K+$^{39}$K$^{23}$Na, $^{87}$Rb+$^{85}$Rb$^{133}$Cs, and $^{88}$Sr+$^{86}$Sr$^{19}$F collisions, respectively. The energy levels of reactants and products are parallel, with different splittings within each $n$ manifold due to the change of hyperfine states. In case of  molecules with lighter atoms, such as NaK or SrF molecules,  the hyperfine splittings of the reactant atoms are much smaller than rotational energy of the molecules. On the other hand, the Rb hyperfine splitting is much larger than the RbCs rotational constant, so the pattern of thresholds of reactants and products is quite complex. Note that the hyperfine states of molecules are superimposed at the resolution of the plot. 

As can be seen in the panels (a) of the figures, it is possible to induce crossings between the energies of reactants and products by tuning the electric field, thus closing or opening reaction thresholds for  the related collisions.  For the $^{39}$K+$^{40}$K$^{23}$Na (Fig.\ref{fig1} (a)), close to 4~kV/cm, the ground state of the reactant  $^{23}$Na$^{39}$K approaches $m_n=0$ states of $^{23}$Na$^{40}$K molecule (which for zero field correlate with $n=1$), and becomes a closed channel while the electric field is increased further. The same situation can be seen close to 7.5~kV$/$cm: the channels with $|m_n| =1$ close as the electric field is swept across the state crossing. 
On the other hand, there are multiple crossings of the ground state of reactants with rotationally excited states of product molecules up to $n=2$ in case of the RbCs system (Fig.\ref{fig2} (a)). A magnitude of electric field above 15~kV/cm closes channels corresponding to $n=2$ states of product  ${}^{87}$Rb${}^{133}$Cs.
The reaction starting with $^{88}$Sr+$^{86}$SrF (see  Fig.\ref{fig3} (a)) is very peculiar, since the reactant ground-state energy is very close to the one of rotationally-excited product states, and an electric field of about 2.5~kV$/$cm is sufficient to close all channels except ground rotational $n=0$ state. The molecules which are produced in $n=1$ state are very slow with total translational energy on the order of 20~mK or less.

 
The statistical approach for collisions was initially formulated in the case of nuclear scattering by Feshbach~\cite{Feshbach:1952,Feshbach:1958,Feshbach:1962}, and adapted for molecular collisions by Bernstein, Light, and Miller~\cite{Bernstein:1963,Light:1965,Miller:1970}. It has been since then extended and combined with more rigorous close-coupling schemes to study the reactive scattering in many-partial wave regime for a number of important chemical reactions~\cite{Rackham:2001,Rackham:2003, Lezana:2006}. Recently, the statistical approach was used for ultracold collision studies between Li and LiYb, and compared to quantum reactive scattering~\cite{Makrides:2015}. Here, the statistical approach of ~\cite{Statfields:2014} is extended to the case of ultracold atom/molecule collisions in a presence of an electric field, and is used to analyze the product state distributions in reactions with exothermicity on the order of tens to hundreds of kelvins. 

An essential assumption of the statistical method is that collisions proceed via the long-lived intermediate complex, so that a full randomization of quantum state populations in the interaction region occurs. Then the reactants and products are uncorrelated, and the transition probability for reactant state $\alpha$ to product state $\beta$ is expressed as 
 \begin{equation}
 \label{stat1}
 P^{M}_{\alpha \to \beta} =  p^M_\alpha p^M_\beta / \sum_\gamma p^M_\gamma,
 \end{equation}
where $\gamma$ denotes all possible values $\beta$. The quantities $p^M_\alpha$ and $p^M_\beta$ are the probabilities for capture in the entrance channel, and isotopologue formation in the outgoing channel. All these probabilities depend on the collision energy $E$ and on the electric field magnitude $\varepsilon$. Note that in the external field the total angular momentum $J$ of the collisional complex is not conserved but its projection $M$ onto the space-fixed axis (along the direction of the electric field) is.

The statistical method should be particularly useful for collisions of heavy atoms with alkali-metal dimers and systems with monofluorides of atoms with an external $s^2$ shell, like alkaline-earth atoms. The method relies on complex formation due to a presence of resonances and for systems with large reduced mass, small rotational constants, and a rich hyperfine structure. The initial estimations of the density of states near the collision threshold were performed by Mayle et al. \cite{Mayle:2013} and recently revisited by Christianen et al. \cite{Christianen:2019} for the systems with neglected hyperfine structure. In the latter paper, the density of states which do not conserve the total angular momentum but conserve its projection (which is the case for electric field) for K$_2$+Rb collisions was calculated to be on the order of $10^4$ states per inverse Kelvin which is two orders of magnitude more than in case when the total angular momentum is conserved (in field-free case). This number is possibly smaller for NaK+K system but larger for RbCs+Rb. The density of states in all systems concerned here is, in fact, more significant due to complex hyperfine structure. The strongest effect which couples different hyperfine states is nuclear quadrupole interaction with the electric field gradients at nucleus which changes with geometry, but also interaction-induced modification of hyperfine coupling of alkali-metal atoms \cite{Barbe:2018} or SrF molecule. The detailed analysis of resonances in these complexes in an electric field with hyperfine couplings included is very challenging. However, for the sake of the present paper, one can safely assume that statistical regime is achieved in all systems studied in the present paper.


The reaction cross-section $\sigma^{M}_{\alpha \rightarrow \beta}$ for a collision at energy $E$ (taken from an arbitrary origin) and an electric field magnitude $\varepsilon$ can be expressed as 
\begin{equation}
\sigma^{M}_{\alpha \rightarrow \beta}(E,\varepsilon) = \frac{\pi \hbar^2 _{}}{2\mu (E-E_{\alpha}(\varepsilon))} \sum_{M} \frac{p^{M}_{\alpha}(E, \varepsilon) p^{M}_{\beta}(E, \varepsilon)}{\sum_{\gamma} p^{M}_{\gamma}(E, \varepsilon)}
\end{equation}
where $\mu$ is the reduced mass of the reactants, $E-E_{\alpha}(\varepsilon)$ gives the effective collisional energy above the reactant energy $E_{\alpha}$ at a given $\varepsilon$.
In the studied cases, the initial channel corresponds to a single hyperfine state of the reactant molecule and of the atom partner, colliding in the $s$-wave regime (\textit{i.e.}, the $\ell$ quantum number associated to the atom-molecule mutual rotation is equal to 0, and it is assumed to be decoupled from the other angular momenta). 
Following ~\cite{Statfields:2014}, we obtained the energy levels of molecules concerned by diagonalizing the  Hamiltonian of the system including the hyperfine structure and the Stark operator (see e.g.~\cite{Aldegunde:2008,Aldegunde:2016}). We used accurate atomic and molecular constants to obtain the energy levels of the systems~\cite{Arimondo:1977,Bize:1999,Childs:1981,Martin:2005, Ross:1990, Ross:2000, Wang:2016, Ferber:2008,Molony:2014,Jastrzebski:2008,Kasahara:1996, Fellows:1999,Deiglmayr:2008,Zuchowski:2013,Brooks:1972,Aldegunde:2008,Aldegunde:2016,Bernath:1996,Ivanova:2011,Ni:2008}. Although the spacing between hyperfine energy levels of $^1\Sigma^+$ molecules is several orders of magnitude smaller than the one between rotational levels, we included them to properly account for state counting. To calculate complex formation probability $P^M_{\alpha,\beta}$, we used  the semi-classical approach of tunneling through the reaction barrier (which in this case reduces to the centrifugal barrier): unit probability for $E-E_{\gamma }$ above the centrifugal barrier (similarly as in Langevin model), and a non-zero probability of the tunneling effect expressed in the Wentzel-Kramers-Brillouin model (WKB) otherwise, which can be written as~\cite{BalintKurtiBook,Statfields:2014}:
\begin{equation}
 p^{M}_{\gamma} (E, \varepsilon) = \exp \Bigg\{ -\frac{2}{\hbar} \int_{R_{\rm min}}^{R_{\rm max}} \sqrt{2 \mu_{\gamma}  \bigg[ \frac{\ell_{\gamma}(\ell_{\gamma}+1) \hbar^2}{2 \mu_{\gamma} R^2 } - \frac{C_6}{R^6} - (E - E_{\gamma}(\varepsilon)) \bigg]  } dR \Bigg\},
\end{equation}
where the $R_{\rm min}$ and $R_{\rm max}$ are classical turning points characterizing a position of the centrifugal barrier and $C_6$ is the long-range dispersion coefficient. We took van der Waals $C_6$ coefficients from Ref.~\cite{Zuchowski:2013} to represent long-range potentials.


Figures \ref{fig1}b, \ref{fig2}b, and \ref{fig3}b show the probabilities of forming the atom+diatom products with the specific translational energy for the reactants with an initial collisional energy of 1 $\mu$K, while the panel (c) therein shows the formation probability of  the molecules in a given $|m_n|$ quantum state. 
Each curve in the panels (c) is a sum of all contributions to the probability presented in the panels (b).  Interestingly,   for zero electric field, the largest propensity of forming the product molecule corresponds to the state with the largest energy difference between ground state $n=0$ and product state. 
That is due to the distribution of products among all accessible $\ell$ quantum numbers in an outgoing channel is the largest.
This result can be rationalized using the phase-space theory (PST) of Pechukas and Light~\cite{Light:1965}, which is essentially a particular case of  Eq.~\ref{stat1} in which step-function probabilities  (equal to 0 for barrier reflection, and equal to 1, for transmission over the barrier) are used. 
If we neglect the hyperfine states, one can easily show that for final rotational state $n$  of products, one should populate $(2n+1)(\ell_{max}(n)+1)/(\sum_{n \leq n_{max}}(2n+1)(\ell_{max}(n) +1) )$, where $\ell_{\rm max}(n)$ is the maximum end-over-end angular momentum for which the centrifugal barrier fits below $\Delta E_{a-b} - B_0 n(n+1)$.  In Table \ref{tbl:zero_field}, we specified the largest possible ${\ell}_{\rm max}$ value corresponding to $n=0$ in the product state. These values are on the order of 20, so clearly, in the outgoing channel, we are no longer in the quantum regime. Approximate estimation of branching ratios in the zero-field with the PST theory agree well with calculations using the theory developed in Ref. \cite{Statfields:2014} for K+NaK and Sr+SrF reactions, for which the energy level splitting due to  hyperfine  coupling is much smaller than the rotational spacing. 

An inspection of Figure \ref{fig1}b reveals that the zero-field probabilities for the K+NaK system are 63\% for the channels corresponding to $n=1$ state with a translational energy order of 100 mK and 37\% for $n=0$ state with energy about 400 mK. Interestingly, product distributions for the K+NaK system are less complicated than in case of the Rb+RbCs reaction(see Figure  \ref{fig2}b). For the latter one, probabilities are 22\% for 60 mK, and 34\% for 110 mK in a mixture of products with $n=1$, and $n=2$ states, and 26\% for separated $n=1$ state with 210 mK, and 9\% for $n=0$ states with energy about 100 mK, and 250 mK. In case of the Sr+SrF reaction, probabilities are 60\% for $n=0$ state with an energy order of 750 mK, and 40\% for energies between about 10 mK, and 20 mK which correspond to $n=1$, and $n=0$ states, respectively.

Reaction products and their probability distributions can be controlled by tuning of an external electric field, since the energy levels have different effective dipole moments (defined as energy derivative with respect to electric field). The Stark shifts of the rotational energy of reactants close reactive channels if product states are above reactant states in the energy. As the energy gap between reactant and product channel changes the number of possible $L_{\rm max}$ states, which can cross the centrifugal barrier decreases. Hence the  number of product states also decreases with the energy gap, as the $L_{\rm max} \approx ( \sqrt{108} \Delta E/E_{\rm vdw})^{\frac{1}{3}}$. For this reason it is rather inefficient to tune the difference between products and reactants arbitrarily close with the electric field as a very small quantity of very cold products will be produced in such reaction. For the $^{40}$K (F=9/2) $+$ ${}^{23}$Na${}^{39}$K (v=0, n=0) reaction, the intensity of the electric field of 8.16 kV/cm is sufficient to produce molecules in the pure rotational ground state. For a slightly detuned value of the electric field, e.g., for the intensity of about 7.3 kV$/$cm, one can produce the molecules in the $|m_n|=1$ state with translational energy of products equaled 16 mK with a probability of 10\%.  Translational energy of products may even be smaller than 100 $\mu$K for about 8.1 kV/cm but with a population of about 3\%. Interestingly, the probability is 23\% for molecules in $n=1$ state with translational energy about 10 $\mu$K in the Rb+RbCs system. In this reaction, the Stark shifts close reactive channels corresponded to $n=1$ state for intensities above 36.3 kV$/$cm. The probability for production of samples of $^{88}$SrF molecules with the translational energy smaller than 100 $\mu$K is 4\% for an electric field  2.6 kV$/$cm which is a slightly detuned value of the field at which all reactive channels close, except ones for the ground rotational state. 


In summary, we have shown that products of the isotopic substitution reactions are cold according to their translational energy below hundreds of mK, and molecular products may occupy the lowest rotational states. We have also discussed the possibility of controlling the chemical reactions by the electric field in ultracold mixtures of molecules and atoms with low kinetic energy release, where one of the constituent atoms of colliding molecule is replaced by its isotope. Such experiments could be implemented with present experimental techniques, using a modest electric field (up to 20 kV$/$cm). With such an experimental setup it is possible to trap a large amount of the product molecules in the trap, if the external microwave cavity trap is used, or - in case of $^2\Sigma^-$ molecules - in a magnetic trap.  This can open new avenues in investigating the branching ratios of chemical reactions in ultracold conditions. Given that full quantum dynamics calculations in external fields are very challenging and at present approximations need to be introduced, such experiments can shed new light on untangling complicated dynamics near the reaction thresholds.

This research has been financed from the funds of the Polish National Science Centre (grants nos. 2017/25/B/ST4/01486 and 2017/27/N/ST4/02576). Calculations have been carried out at the Wroclaw Centre for Networking and Supercomputing (http://www.wcss.pl), Grant No. 218 (PSŻ). This work is part of the Polonium programme.

\section*{References}
\renewcommand{\section}[2]{}%

\bibliography{refer_mk}

\providecommand{\newblock}{}
\begin{thebibliography}{10}
\expandafter\ifx\csname url\endcsname\relax
  \def\url#1{{\tt #1}}\fi
\expandafter\ifx\csname urlprefix\endcsname\relax\def\urlprefix{URL }\fi
\providecommand{\eprint}[2][]{\url{#2}}

\bibitem{Ospelkaus:2009}
Ospelkaus S, Ni K~K, Wang D, de~Miranda M~H~G, Neyenhuis B, Qu{\'e}m{\'e}ner G,
  Julienne P~S, Bohn J~L, Jin D~S and Ye J 2010 {\em Science\/} {\bf 327}
  853--857 \urlprefix\url{http://science.sciencemag.org/content/327/5967/853}

\bibitem{Ni:2010}
Ni K~K, Ospelkaus S, Wang D, Qu{\'e}m{\'e}ner G, Neyenhuis B, de~Miranda M~H~G,
  Bohn J~L, J Y and Jin D 2010 {\em Nature\/} {\bf 464} 1324
  \urlprefix\url{http://dx.doi.org/10.1038/nature08953}

\bibitem{deMiranda:2011}
de~Miranda M~H~G, Chotia A, Neyenhuis B, Wang D, Qu\'em\'ener G, Ospelkaus S,
  Bohn J~L, Ye J and Jin D~S 2011 {\em Nat. Phys.\/} {\bf 7} 502--507
  \urlprefix\url{http://dx.doi.org/10.1038/nphys1939}

\bibitem{Ye:2018}
Ye X, Guo M, Gonz{\'a}lez-Mart{\'\i}nez M~L, Qu{\'e}m{\'e}ner G and Wang D 2018
  {\em Science Advances\/} {\bf 4}
  \urlprefix\url{http://advances.sciencemag.org/content/4/1/eaaq0083}

\bibitem{Knoop:2010}
Knoop S, Ferlaino F, Berninger M, Mark M, N\"agerl H~C, Grimm R, D'Incao J~P
  and Esry B~D 2010 {\em Phys. Rev. Lett.\/} {\bf 104}(5) 053201
  \urlprefix\url{https://link.aps.org/doi/10.1103/PhysRevLett.104.053201}

\bibitem{Rui:2017}
Rui J, Yang H, Liu L, Zhang D~C, Liu Y~X, Nan J, Chen Y~A, Zhao B and Pan J~W
  2017 {\em Nat. Phys.\/} {\bf 13} 699--703
  \urlprefix\url{http://dx.doi.org/10.1038/nphys4095}

\bibitem{Zuchowski:2010}
\.Zuchowski P~S and Hutson J~M 2010 {\em Phys. Rev. A\/} {\bf 81} 060703
  \urlprefix\url{https://link.aps.org/doi/10.1103/PhysRevA.81.060703}

\bibitem{Kosicki:2017}
Kosicki M~B, , K\ifmmode~\mbox{\c{e}}\else \c{e}\fi{}dziera D and
  \ifmmode~\dot{Z}\else \.{Z}\fi{}uchowski P~S 2017 {\em J. Phys. Chem. A\/}
  {\bf 121} 4152--4159
  \urlprefix\url{http://dx.doi.org/10.1021/acs.jpca.7b01523}

\bibitem{Zwierlein:2015}
Park J~W, Will S~A and Zwierlein M~W 2015 {\em Phys. Rev. Lett.\/} {\bf 114}
  205302 \urlprefix\url{http://link.aps.org/doi/10.1103/PhysRevLett.114.205302}

\bibitem{Will:2016}
Will S~A, Park J~W, Yan Z~Z, Loh H and Zwierlein M~W 2016 {\em Phys. Rev.
  Lett.\/} {\bf 116}(22) 225306
  \urlprefix\url{https://link.aps.org/doi/10.1103/PhysRevLett.116.225306}

\bibitem{Wang:2016}
Guo M, Zhu B, Lu B, Ye X, Wang F, Vexiau R, Bouloufa-Maafa N, Qu\'em\'ener G,
  Dulieu O and Wang D 2016 {\em Phys. Rev. Lett.\/} {\bf 116}(20) 205303
  \urlprefix\url{http://link.aps.org/doi/10.1103/PhysRevLett.116.205303}

\bibitem{Cornish:2014}
Molony P~K, Gregory P~D, Ji Z, Lu B, K\"oppinger M~P, Le~Sueur C~R, Blackley
  C~L, Hutson J~M and Cornish S~L 2014 {\em Phys. Rev. Lett.\/} {\bf 113}
  255301 \urlprefix\url{http://link.aps.org/doi/10.1103/PhysRevLett.113.255301}

\bibitem{Takekoshi:2014}
Takekoshi T, Reichs\"{o}llner L, Schindewolf A, Hutson J~M, Le~Sueur C~R,
  Dulieu O, Ferlaino F, Grimm R and N\"{a}gerl H~C 2014 {\em Phys. Rev.
  Lett.\/} {\bf 113} 205301
  \urlprefix\url{http://link.aps.org/doi/10.1103/PhysRevLett.113.205301}

\bibitem{Truppe:2017}
Truppe S, Williams H~J, Hambach M, Caldwell L, Fitch N~J, Hinds E~A, Sauer B~E
  and Tarbutt M~R 2017 {\em Nat. Phys.\/} {\bf 13} 1173–1176
  \urlprefix\url{https://www.nature.com/articles/nphys4241}

\bibitem{Williams:2018}
Williams H~J, Caldwell L, Fitch N~J, Truppe S, Rodewald J, Hinds E~A, Sauer B~E
  and Tarbutt M~R 2018 {\em Phys. Rev. Lett.\/} {\bf 120}(16) 163201
  \urlprefix\url{https://link.aps.org/doi/10.1103/PhysRevLett.120.163201}

\bibitem{McCarron:2018}
McCarron D~J, Steinecker M~H, Zhu Y and DeMille D 2018 {\em Phys. Rev. Lett.\/}
  {\bf 121}(1) 013202
  \urlprefix\url{https://link.aps.org/doi/10.1103/PhysRevLett.121.013202}

\bibitem{Anderegg:2018}
Anderegg L, Augenbraun B~L, Bao Y, Burchesky S, Cheuk L~W, Ketterle W and Doyle
  J~M 2018 {\em Nat. Phys.\/} {\bf 14} 890–893
  \urlprefix\url{https://doi.org/10.1038/s41567-018-0191-z}

\bibitem{Borsalino:2016}
Borsalino D, Vexiau R, Aymar M, Luc-Koenig E, Dulieu O and Bouloufa-Maafa N
  2016 {\em J. Phys. B: At. Mol. Opt. Phys.\/} {\bf 49} 055301
  \urlprefix\url{http://stacks.iop.org/0953-4075/49/i=5/a=055301}

\bibitem{Stevenson:2016}
Stevenson I~C, Blasing D~B, Chen Y~P and Elliott D~S 2016 {\em Phys. Rev. A\/}
  {\bf 94} 062510
  \urlprefix\url{https://link.aps.org/doi/10.1103/PhysRevA.94.062510}

\bibitem{Harter:2013}
H\"{a}rter A, Kr\"{u}kow A, Dei{\ss} M, Drews B, Tiemann E and Denschlag J~H
  2013 {\em Nat. Phys.\/} {\bf 9} 512--517
  \urlprefix\url{http://dx.doi.org/10.1038/nphys2661}

\bibitem{Wolf:2017}
Wolf J, Dei{\ss} M, Kr\"{u}kow A, Tiemann E, Ruzic B~P, Wang Y, D'Incao J~P,
  Julienne P~S and Denschlag J~H 2017 {\em Science\/} {\bf 358} 921--924
  \urlprefix\url{http://science.sciencemag.org/content/358/6365/921}

\bibitem{Hu:2019}
Hu M~G, Liu Y, Grimes D~D, Lin Y~W, Gheorghe A~H, Vexiau R, Boulufa-Maafa N,
  Dulieu O and Ni K~K 2019 {\em arXiv:1907.13628\/}

\bibitem{Tomza:2015:b}
Tomza M 2015 {\em Phys. Rev. Lett.\/} {\bf 115} 063201

\bibitem{Hutson:2006}
Hutson J~M and Sold\'an P 2007 {\em Int. Rev. Phys. Chem.\/} {\bf 26} 1--28
  \urlprefix\url{https://doi.org/10.1080/01442350601084562}

\bibitem{Cvitas:2005a}
Cvita\v{s} M~T, Sold\'an P, Hutson J~M, Honvault P and Launay J~M 2005 {\em
  Phys. Rev. Lett.\/} {\bf 94}(3) 033201
  \urlprefix\url{https://link.aps.org/doi/10.1103/PhysRevLett.94.033201}

\bibitem{Cvitas:2005b}
Cvita\v{s} M~T, Sold\'an P, Hutson J~M, Honvault P and Launay J~M 2005 {\em
  Phys. Rev. Lett.\/} {\bf 94}(20) 200402
  \urlprefix\url{https://link.aps.org/doi/10.1103/PhysRevLett.94.200402}

\bibitem{Cvitas:2007}
Cvita\v{s} M~T, Sold\'{a}n P, Hutson J~M, Honvault P and Launay J~M 2007 {\em
  J. Chem. Phys\/} {\bf 127} 074302
  \urlprefix\url{http://dx.doi.org/10.1063/1.2752162}

\bibitem{Quemener:2005}
Qu\'em\'ener G, Honvault P, Launay J~M, Sold\'an P, Potter D~E and Hutson J~M
  2005 {\em Phys. Rev. A\/} {\bf 71}(3) 032722
  \urlprefix\url{https://link.aps.org/doi/10.1103/PhysRevA.71.032722}

\bibitem{Croft:2017a}
Croft J~F~E, Makrides C, Li M, Petrov A, K K~B, Balakrishnan N and Kotochigova
  S 2017 Universality and chaoticity in ultracold k+krb chemical reactions
  \urlprefix\url{http://dx.doi.org/10.1038/ncomms15897}

\bibitem{Croft:2017b}
Croft J~F~E, Balakrishnan N and Kendrick B~K 2017 {\em Phys. Rev. A\/} {\bf
  96}(6) 062707
  \urlprefix\url{https://link.aps.org/doi/10.1103/PhysRevA.96.062707}

\bibitem{Lara:2015a}
Lara M, Jambrina P~G, Launay J~M and Aoiz F~J 2015 {\em Phys. Rev. A\/} {\bf
  91}(3) 030701
  \urlprefix\url{https://link.aps.org/doi/10.1103/PhysRevA.91.030701}

\bibitem{Lara:2015b}
Lara M, Jambrina P~G, Aoiz F~J and Launay J~M 2015 {\em J. Chem. Phys.\/} {\bf
  143} 204305 \urlprefix\url{https://doi.org/10.1063/1.4936144}

\bibitem{Mayle:2012}
Mayle M, Ruzic B~P and Bohn J~L 2012 {\em Phys. Rev. A\/} {\bf 85}(6) 062712
  \urlprefix\url{https://link.aps.org/doi/10.1103/PhysRevA.85.062712}

\bibitem{Statfields:2014}
Gonz\'alez-Mart\'{\i}nez M~L, Dulieu O, Larr\'egaray P and Bonnet L 2014 {\em
  Phys. Rev. A\/} {\bf 90}(5) 052716
  \urlprefix\url{https://link.aps.org/doi/10.1103/PhysRevA.90.052716}

\bibitem{Ivanova:2011}
Ivanova M, Stein A, Pashov A, Kn\"ockel H and Tiemann E 2011 {\em J. Chem.
  Phys.\/} {\bf 134} 024321 \urlprefix\url{https://doi.org/10.1063/1.3524312}

\bibitem{Jastrzebski:2008}
Adohi-Krou A, Jastrzebski W, Kowalczyk P, Stolyarov A and Ross A 2008 {\em J.
  Mol. Spectrosc.\/} {\bf 250} 27--32
  \urlprefix\url{http://www.sciencedirect.com/science/article/pii/S0022285208001446}

\bibitem{Ross:1990}
Ross A~J, Effantin C, Crozet P and Boursey E 1990 {\em J. Phys. B: At. Mol.
  Opt. Phys.\/} {\bf 23} L247
  \urlprefix\url{http://stacks.iop.org/0953-4075/23/i=12/a=002}

\bibitem{Ferber:2008}
Ferber R, Klincare I, Nikolayeva O, Tamanis M, Kn\"{o}ckel H, Tiemann E and
  Pashov A 2008 {\em J. Chem. Phys.\/} {\bf 128} 244316
  \urlprefix\url{http://dx.doi.org/10.1063/1.2943677}

\bibitem{Kasahara:1996}
Kasahara S, Ebi T, Tanimura M, Ikoma H, Matsubara K, Baba M and Kat\^{o} H 1996
  {\em J. Chem. Phys.\/} {\bf 105} 1341--1347
  \urlprefix\url{http://dx.doi.org/10.1063/1.472000}

\bibitem{Fellows:1999}
Fellows C~E, Gutterres R~F, Campos A~P~C, Verg\`{e}s J and Amiot C 1999 {\em J.
  Mol. Spectrosc.\/} {\bf 197} 19--27
  \urlprefix\url{http://www.sciencedirect.com/science/article/pii/S0022285299978803}

\bibitem{Bernath:1996}
Colarusso P, Guo B, Zhang K~Q,  and Bernath D 1996 {\em J. Mol. Spectr.\/} {\bf
  175} 158--171

\bibitem{Lutz:2016}
Lutz J~J and Hutson J~M 2016 {\em J. Mol. Spectrosc.\/} {\bf 330} 43--56
  \urlprefix\url{https://doi.org/10.1016/j.jms.2016.08.007}

\bibitem{ADF2001}
te~Velde G, Bickelhaupt F~M, Baerends E~J, Fonseca~Guerra C, van Gisbergen
  S~J~A, Snijders J~G and Ziegler T 2001 {\em J. Comput. Chem.\/} {\bf 22}
  931--967 ISSN 1096-987X \urlprefix\url{http://dx.doi.org/10.1002/jcc.1056}

\bibitem{PBE96}
Perdew J~P, Burke K and Ernzerhof M 1996 {\em Phys. Rev. Lett.\/} {\bf 77}(18)
  3865--3868
  \urlprefix\url{https://link.aps.org/doi/10.1103/PhysRevLett.77.3865}

\bibitem{Ross:2000}
Russier-Antoine I, Ross A, Aubert-Fr\'{e}con M, Martin F and Crozet P 2000 {\em
  J. Phys. B: At. Mol. Opt. Phys.\/} {\bf 33} 2753–2762
  \urlprefix\url{http://stacks.iop.org/0953-4075/33/i=14/a=312}

\bibitem{Arimondo:1977}
Arimondo E, Inguscio M and Violino P 1977 {\em Rev. Mod. Phys.\/} {\bf 49}(1)
  31--75 \urlprefix\url{https://link.aps.org/doi/10.1103/RevModPhys.49.31}

\bibitem{Ni:2008}
Ni K~K, Ospelkaus S, de~Miranda M~H~G, Peer A, Neyenhuis B, Zirbel J~J,
  Kotochigova S, Julienne P~S, Jin D~S and Ye J 2008 {\em Science\/} {\bf 322}
  231--235 \urlprefix\url{http://science.sciencemag.org/content/322/5899/231}

\bibitem{Deiglmayr:2008}
Deiglmayr J, Aymar M, Wester R, Weidem\"{u}ller M and Dulieu O 2008 {\em J.
  Chem. Phys.\/} {\bf 129} 064309
  \urlprefix\url{http://dx.doi.org/10.1063/1.2960624}

\bibitem{Bize:1999}
Bize S, Sortais Y, Santos M~S, Mandache C, Clairon A and Salomon C 1999 {\em
  Europhys. Lett.\/} {\bf 45} 558
  \urlprefix\url{http://stacks.iop.org/0295-5075/45/i=5/a=558}

\bibitem{Childs:1981}
Childs W~J, Goodman G~L and Goodman L~S 1981 {\em J. Mol. Spectrosc.\/} {\bf
  86} 365--392 \urlprefix\url{https://doi.org/10.1016/0022-2852(81)90288-5}

\bibitem{Aymar:2005}
Aymar M and Dulieu O 2005 {\em J. Chem. Phys.\/} {\bf 122} 204302

\bibitem{Feshbach:1952}
Hauser W and Feshbach H 1952 {\em Phys. Rev.\/} {\bf 87}(2) 366--373
  \urlprefix\url{https://link.aps.org/doi/10.1103/PhysRev.87.366}

\bibitem{Feshbach:1958}
Feshbach H 1958 {\em Ann. Phys.\/} {\bf 5} 357
  \urlprefix\url{https://doi.org/10.1016/0003-4916(58)90007-1}

\bibitem{Feshbach:1962}
Feshbach H 1962 {\em Ann. Phys.\/} {\bf 19} 287
  \urlprefix\url{https://doi.org/10.1016/0003-4916(62)90221-X}

\bibitem{Bernstein:1963}
Bernstein R~B, Dalgarno A, Massey H and Percival I~C 1963 {\em Proc. R. Soc.
  Lond. A\/} {\bf 274} 427--442
  \urlprefix\url{https://doi.org/10.1098/rspa.1963.0142}

\bibitem{Light:1965}
Pechukas P and Light J~C 1965 {\em J. Chem. Phys.\/} {\bf 42} 3281
  \urlprefix\url{http://dx.doi.org/10.1063/1.1696411}

\bibitem{Miller:1970}
Miller W~H 1970 {\em J. Chem. Phys.\/} {\bf 52} 543
  \urlprefix\url{http://dx.doi.org/10.1063/1.1673020}

\bibitem{Rackham:2001}
Rackham E~J, Huarte-Larranaga F and Manolopoulos D~E 2001 {\em Chem. Phys.
  Lett.\/} {\bf 343} 356--364
  \urlprefix\url{http://dx.doi.org/10.1016/S0009-2614(01)00707-2}

\bibitem{Rackham:2003}
Rackham E~J, Gonz\'{a}lez-Lezana T and Manolopoulos D~E 2003 {\em J. Chem.
  Phys.\/} {\bf 119} 12895--12907
  \urlprefix\url{http://dx.doi.org/10.1063/1.1628218}

\bibitem{Lezana:2006}
Gonz\'{a}lez-Lezana T 2006 {\em Int. Rev. Phys. Chem.\/} {\bf 26} 29--91
  \urlprefix\url{http://dx.doi.org/10.1080/03081070600933476}

\bibitem{Makrides:2015}
Makrides C, Hazra J, Pradhan G~B, Petrov A, Kendrick B~K, Gonz\'alez-Lezana T,
  Balakrishnan N and Kotochigova S 2015 {\em Phys. Rev. A\/} {\bf 91}(1) 012708
  \urlprefix\url{https://link.aps.org/doi/10.1103/PhysRevA.91.012708}

\bibitem{Mayle:2013}
Mayle M, Qu\'em\'ener G, Ruzic B~P and Bohn J~L 2013 {\em Phys. Rev. A\/} {\bf
  87}(1) 012709
  \urlprefix\url{https://link.aps.org/doi/10.1103/PhysRevA.87.012709}

\bibitem{Christianen:2019}
Christianen A, Karman T and Groenenboom G~C 2019 {\em arXiv:1905.06691v3\/}

\bibitem{Barbe:2018}
Barb\'e V, Ciamei A, Pasquiou B, Reichs\"ollner L, Schreck F, \.Zuchowski P~S
  and Hutson J~M 2018 {\em Nat. Phys.\/} {\bf 14} 881–884
  \urlprefix\url{https://doi.org/10.1038/s41567-018-0169-x}

\bibitem{Aldegunde:2008}
Aldegunde J, Rivington B~A, \ifmmode~\dot{Z}\else \.{Z}\fi{}uchowski P~S and
  Hutson J~M 2008 {\em Phys. Rev. A\/} {\bf 78}(3) 033434
  \urlprefix\url{https://link.aps.org/doi/10.1103/PhysRevA.78.033434}

\bibitem{Aldegunde:2016}
Gregory P~D, Aldegunde J, Hutson J~M and Cornish S~L 2016 {\em Phys. Rev. A\/}
  {\bf 94}(4) 041403
  \urlprefix\url{https://link.aps.org/doi/10.1103/PhysRevA.94.041403}

\bibitem{Martin:2005}
Sansonetti J~E and Martin W~C 2005 {\em J. Phys. Chem. Ref. Data\/} {\bf 34}
  1559

\bibitem{Molony:2014}
Molony P~K, Gregory P~D, Ji Z, Lu B, K\"oppinger M~P, Le~Sueur C~R, Blackley
  C~L, Hutson J~M and Cornish S~L 2014 {\em Phys. Rev. Lett.\/} {\bf 113}(25)
  255301
  \urlprefix\url{https://link.aps.org/doi/10.1103/PhysRevLett.113.255301}

\bibitem{Zuchowski:2013}
\ifmmode~\dot{Z}\else \.{Z}\fi{}uchowski P~S, Kosicki M, Kodrycka M and
  Sold\'an P 2013 {\em Phys. Rev. A\/} {\bf 87}(2) 022706
  \urlprefix\url{https://link.aps.org/doi/10.1103/PhysRevA.87.022706}

\bibitem{Brooks:1972}
Brooks R~A, Anderson C~H and Ramsey N~F 1972 {\em J. Chem. Phys.\/} {\bf 56}
  5193 \urlprefix\url{http://dx.doi.org/10.1063/1.1677011}

\bibitem{BalintKurtiBook}
Balint-Kurti G~G and Palov A~P 2015 {\em Theory of Molecular Collisions\/} (The
  Royal Society of Chemistry) ISBN 978-1-78262-019-8

\bibitem{Irikura:2007}
Irikura K~K 2007 {\em J. Phys. Chem. Ref. Data,\/} {\bf 36} 389
  \urlprefix\url{https://doi.org/10.1063/1.2436891}

\bibitem{Hessel:1971}
Hessel M~M 1971 {\em Phys. Rev. Lett.\/} {\bf 26}(5) 215--218
  \urlprefix\url{https://link.aps.org/doi/10.1103/PhysRevLett.26.215}

\bibitem{Bednarska:1998}
Bednarska V, Jackowska I, Jastrz\ifmmode~\mbox{\c{e}}\else \c{e}\fi{}bski W and
  Kowalczyk P 1998 {\em J. Mol. Spectrosc.\/} {\bf 189} 244--248
  \urlprefix\url{https://doi.org/10.1006/jmsp.1998.7543}

\bibitem{Ridinger:2011}
Ridinger A, Chaudhuri S, Salez T, Fernandes D~R, Bouloufa N, Dulieu O, Salomon
  C and Chevy F 2011 {\em Europhys. Lett.\/} {\bf 96} 33001
  \urlprefix\url{https://doi.org/10.1209/0295-5075/96/33001}

\bibitem{Staanum:2007}
Staanum P, Pashov A, Kn\"ockel H and Tiemann E 2007 {\em Phys. Rev. A\/} {\bf
  75}(4) 042513
  \urlprefix\url{https://link.aps.org/doi/10.1103/PhysRevA.75.042513}

\end{thebibliography}

\newpage

\subsection{Supplementary material}

As mentioned in Table \ref{tbl:zero_field}, we present complete list of the kinetic energy release, $\Delta E_{a-b}$, for all possible isotopic substitution reactions in the alkali-metal dimers. These values are obtained using the Eq.~\ref{eq:deltaE}.

\begin{table*}[ht]
    \label{tab2}
\caption{The calculated reaction energy $\Delta E_{a-b}$ (references to used data in square brackets),  and the rotational constant $B_0$, the hyperfine coupling constant $A$ of products are shown. Symbols $n_{\rm max}$ and ${\ell}_{\rm max}$ denote the maximum  
rotational state of molecule for a collision energy of 1 $\mu$K and  the maximum end-over-end angular momentum which can be populated by $\Delta E_{a-b}$, respectively.}
\centering
\begin{tabular}{p{0.2\linewidth}p{0.14\linewidth}p{0.15\linewidth}p{0.14\linewidth}p{0.12\linewidth}p{0.05\linewidth}p{0.05\linewidth}p{0.05\linewidth}}
\hline
Products & $\Delta E_{a-b}$  $[$mK$]$ &   $B_0/{k_b T}$ $[$mK$]$ & $A/{k_b T}$ $[$mK$]$& $n_{\rm max}$ &  ${\ell}_{\rm max}$\\
\hline
${}^{6}$Li $+$ ${}^{7}$Li${}^{23}$Na& 10747~\cite{Irikura:2007}&  570~\cite{Hessel:1971}   &   73~\cite{Arimondo:1977} & 1 & 21   \\
${}^{6}$Li $+$ ${}^{7}$Li${}^{39}$K &   974~\cite{Bednarska:1998} & 379~\cite{Ridinger:2011} & 73~\cite{Arimondo:1977} & 1 & 22   \\
${}^{6}$Li $+$ ${}^{7}$Li${}^{40}$K &   975~\cite{Bednarska:1998} & 379~\cite{Ridinger:2011} & 73~\cite{Arimondo:1977} & 1 & 22   \\
${}^{6}$Li $+$ ${}^{7}$Li${}^{41}$K &   977~\cite{Bednarska:1998} & 379~\cite{Ridinger:2011} & 73~\cite{Arimondo:1977} & 1 & 22  \\
${}^{6}$Li $+$ ${}^{7}$Li${}^{85}$Rb &  969~\cite{Ivanova:2011} &   316~\cite{Deiglmayr:2008} & 73~\cite{Arimondo:1977} & 1 & 23      \\
${}^{6}$Li $+$ ${}^{7}$Li${}^{87}$Rb &  969~\cite{Ivanova:2011} &   316~\cite{Deiglmayr:2008} & 73~\cite{Arimondo:1977} & 1 & 23   \\
${}^{6}$Li $+$ ${}^{7}$Li${}^{133}$Cs & 940~\cite{Staanum:2007} &   279~\cite{Deiglmayr:2008} & 73~\cite{Arimondo:1977} & 1 & 24   \\
${}^{39}$K $+$ ${}^{6}$Li${}^{40}$K & 273~\cite{Bednarska:1998} & 379~\cite{Ridinger:2011} & 11~\cite{Arimondo:1977} & 0 & 14  \\
${}^{39}$K $+$ ${}^{6}$Li${}^{41}$K & 532~\cite{Bednarska:1998} & 379~\cite{Ridinger:2011} & 11~\cite{Arimondo:1977} & 0 & 18  \\
${}^{40}$K $+$ ${}^{6}$Li${}^{41}$K & 259~\cite{Bednarska:1998} & 379~\cite{Ridinger:2011} & -14~\cite{Arimondo:1977} & 0 & 15  \\
${}^{39}$K $+$ ${}^{7}$Li${}^{40}$K & 291~\cite{Bednarska:1998} & 379~\cite{Ridinger:2011} & 11~\cite{Arimondo:1977} &  0 & 15  \\
${}^{39}$K $+$ ${}^{7}$Li${}^{41}$K & 569~\cite{Bednarska:1998} & 379~\cite{Ridinger:2011} & 11~\cite{Arimondo:1977} &  0 & 18  \\
${}^{40}$K $+$ ${}^{7}$Li${}^{41}$K & 277~\cite{Bednarska:1998} & 379~\cite{Ridinger:2011} & -14~\cite{Arimondo:1977} & 0 & 14  \\
${}^{39}$K $+$ ${}^{23}$Na${}^{40}$K    &  415~\cite{Jastrzebski:2008}  & 136~\cite{Ross:2000}	 	& 11~\cite{Arimondo:1977}  &  1 & 18 \\
${}^{39}$K $+$ ${}^{23}$Na${}^{41}$K    & 811~\cite{Jastrzebski:2008} 	&  136~\cite{Ross:2000} 	& 11~\cite{Arimondo:1977}  & 1 & 23\\
${}^{40}$K $+$ ${}^{23}$Na${}^{41}$K    & 396~\cite{Jastrzebski:2008} 	& 136~\cite{Ross:2000} 		& -14~\cite{Arimondo:1977} &   1 & 18\\
${}^{39}$K $+$ ${}^{40}$K${}^{85}$Rb &470~\cite{Ross:1990} &  55~\cite{Ni:2008} & 11~\cite{Arimondo:1977} &    2 & 22   \\
${}^{39}$K $+$ ${}^{41}$K${}^{85}$Rb &920~\cite{Ross:1990} &  55~\cite{Ni:2008} & 11~\cite{Arimondo:1977} &  3 & 26   \\
${}^{40}$K $+$ ${}^{41}$K${}^{85}$Rb &450~\cite{Ross:1990} &  55~\cite{Ni:2008} & -14~\cite{Arimondo:1977} & 2  & 22 \\
${}^{39}$K $+$ ${}^{40}$K${}^{87}$Rb    & 472~\cite{Ross:1990} 			& 55~\cite{Ni:2008} 		& 11~\cite{Arimondo:1977} & 2 & 22 \\
${}^{39}$K $+$ ${}^{41}$K${}^{87}$Rb &923~\cite{Ross:1990} & 55~\cite{Ni:2008} & 11~\cite{Arimondo:1977} &   3 &28  \\
${}^{40}$K $+$ ${}^{41}$K${}^{87}$Rb &451~\cite{Ross:1990} & 55~\cite{Ni:2008} & -14~\cite{Arimondo:1977} &  2  & 22   \\
${}^{39}$K $+$ ${}^{40}$K${}^{133}$Cs   & 478~\cite{Ferber:2008} & 44~\cite{Deiglmayr:2008} 	& 11~\cite{Arimondo:1977} &  2 & 23 \\
${}^{39}$K $+$ ${}^{41}$K${}^{133}$Cs &937~\cite{Ferber:2008} & 44~\cite{Deiglmayr:2008}  & 11~\cite{Arimondo:1977} & 4   & 29  \\
${}^{40}$K $+$ ${}^{41}$K${}^{133}$Cs &458~\cite{Ferber:2008} & 44~\cite{Deiglmayr:2008}  & -14~\cite{Arimondo:1977} &  2  & 23  \\
${}^{85}$Rb $+$ ${}^{6}$Li${}^{87}$Rb & 115~\cite{Ivanova:2011} &   316~\cite{Deiglmayr:2008} &  49~\cite{Bize:1999} & 0  & 16 \\
${}^{85}$Rb $+$ ${}^{7}$Li${}^{87}$Rb   & 123~\cite{Ivanova:2011} 		&   316~\cite{Deiglmayr:2008} & 49~\cite{Bize:1999}     &  0 & 16 \\
${}^{85}$Rb $+$ ${}^{23}$Na${}^{87}$Rb  & 188~\cite{Kasahara:1996} 		&  100~\cite{Wang:2016} 	& 49~\cite{Bize:1999}     & 0 & 20 \\
${}^{85}$Rb $+$ ${}^{39}$K${}^{87}$Rb   & 198~\cite{Ross:1990}			&  55~\cite{Ni:2008} 		& 49~\cite{Bize:1999}     & 1 & 22 \\
${}^{85}$Rb $+$ ${}^{40}$K${}^{87}$Rb &199~\cite{Ross:1990} &   55~\cite{Ni:2008} & 49~\cite{Bize:1999}     & 1  & 22 \\
${}^{85}$Rb $+$ ${}^{41}$K${}^{87}$Rb &201~\cite{Ross:1990} &  55~\cite{Ni:2008} & 49~\cite{Bize:1999}     &1  & 22 \\ 
${}^{85}$Rb $+$ ${}^{87}$Rb${}^{133}$Cs & 253~\cite{Fellows:1999}	 	& 23~\cite{Childs:1981} 	& 49~\cite{Bize:1999}     &  1 & 27 \\
\hline

\end{tabular}
\end{table*}

\end{document}